\documentclass[epj]{svjour}
\usepackage{graphics} 
\usepackage{graphicx}% Include figure files
\usepackage{dcolumn}% Align table columns on decimal point
\usepackage{bm}% bold math
\usepackage{cite}
\usepackage{esvect}
\usepackage{booktabs}
\usepackage{subfigure}
\usepackage{multirow}
\usepackage{float} 

\begin{document}
\title{Universal scaling of strange particle $p_{\rm T}$ spectra in pp collisions}
\author{Liwen Yang, Yanyun Wang, Wenhui Hao, Na Liu, Xiaoling Du, \and Wenchao Zhang% etc
%\thanks{\emph{wenchao.zhang@snnu.edu.cn} Insert the address here if needed}%
}                     % Do not remove
\institute{School of Physics and Information Technology, Shaanxi Normal University, Xi'an 710119, People's Republic of China \\
  \email{wenchao.zhang@snnu.edu.cn}}
\date{Received: date / Revised version: date}
\abstract{
As a complementary study to that performed on the transverse momentum ($p_{\rm T}$) spectra of charged pions, kaons and protons in proton-proton (pp) collisions at LHC energies 0.9, 2.76 and 7 TeV, we present a scaling behaviour in the $p_{\rm T}$ spectra of strange particles ($K_{S}^{0}$, $\rm \Lambda$, $\rm \Xi$ and $\phi$)  at these three energies. This scaling behaviour is exhibited when the spectra are expressed in a suitable scaling variable $z=p_{\rm T}/K$, where the scaling parameter $K$ is determined by the quality factor method and increases with the center of mass energy ($\sqrt{s}$). The rates at which $K$ increases with $\mathrm{ln}\sqrt{s}$ for these strange particles are found to be identical within errors. In the framework of the colour string percolation model, we argue that these strange particles are produced through the decay of clusters that are formed by the colour strings overlapping. We observe that the strange mesons and baryons are produced from clusters with different size distributions, while the strange mesons (baryons) $K_{S}^{0}$ and $\phi$ ($\rm \Lambda$ and $\rm \Xi$) originate from clusters with the same size distributions. The cluster's size distributions for strange mesons are more dispersed than those for strange baryons.  The scaling behaviour of the $p_{\rm T}$ spectra for these strange particles can be explained by the colour string percolation model in a quantitative way.
%We observed that the dispersions of the cluster's size distribution for the strange mesons $K_{S}^{0}$ and $\phi$ are larger than those for the strange baryons $\rm \Lambda$ and $\rm \Xi$, while the dispersion of the cluster's size distribution for $K_{S}^{0}$ ($\rm \Lambda$) is identical to that for $\phi$ ($\rm \Xi$).  This implies that t
 % end of PACS codes
} %end of abstract
\authorrunning{Liwen Yang, et al.}
\titlerunning{Universal scaling of strange particle $p_{\rm T}$ spectra in pp collisions}
\maketitle
\section{Introduction}
\label{sec:intro}
The transverse momentum ($p_{\rm T}$) spectra of final state particles are important observables in high energy collisions. They play an essential role in understanding the mechanism of particle productions. In many studies, searching for a scaling behaviour of the $p_{\rm T}$ spectra is useful to reveal the mechanism. In ref. \cite{pion_spectrum}, a scaling behaviour was presented in the pion $p_{\rm T}$ spectra in Au-Au collisions at the Relativistic Heavy Ion Collider (RHIC). It was independent of the centrality of the collision. This scaling behaviour was later extended to the proton and anti-proton $p_{\rm T}$ spectra with different centralities in Au-Au collisions at RHIC \cite{proton_antiproton_spectra}.

Recently, a similar scaling behaviour was found in the $p_{\rm T}$ spectra of inclusive charged hadrons as well as identified charged hadrons (charged pions, kaons and protons) in proton-proton (pp) collisions at the Large Hadron Collider (LHC) \cite{inclusive_scaling, pi_k_p_scaling}. This scaling behaviour was independent of the center of mass energy ($\sqrt{s}$). It was exhibited when the spectra were expressed in a suitable scaling variable $z=p_{\rm T}/K$, where $K$ is the scaling parameter relying on $\sqrt{s}$. In pp collisions, the hadrons produced are predominantly pions, kaons and protons. As the strange quark is heavier than the up and down quarks, the strange particles such as $K_{S}^{0}$, $\rm \Lambda$, $\rm \Xi$ and $\phi$ only constitute a small fraction of final state particles. However, the investigation of their spectra is an important ingredient in understanding the mechanism of particle production in high energy collisions. Thus, in this paper, we will focus on the $p_{\rm T}$ spectra of $K_{S}^{0}$, $\rm \Lambda$, $\rm \Xi$ and $\phi$ produced in pp collisions at 0.9, 2.76 and 7 TeV \cite{strange_production_1, strange_production_2, strange_production_3, strange_production_4, strange_production_5, strange_production_6}. The $p_{\rm T}$ spectra of $\rm \Omega$ are not considered in this work, as their spectra at 0.9 TeV are not available so far. A scaling behaviour independent of the collision energy will be searched for among these strange particle spectra. If the scaling behaviour exists, then one may ask two questions: (1) Is the dependence of the scaling parameter $K$ on $\sqrt{s}$ for $K_{S}^{0}$, $\rm \Lambda$, $\rm \Xi$ and $\phi$ the same as that for charged pions, kaons and protons? (2) Can the string percolation model utilized in ref. \cite{pi_k_p_scaling} be adopted to explain the scaling behaviour of strange particles?

The organization of the paper is as follows. In sect. \ref{sec:method}, the method to search for the scaling behaviour will be described briefly. In sect. \ref{sec:scaling_behaviour}, the scaling behaviour of the $K_{S}^{0}$, $\rm \Lambda$, $\rm \Xi$ and $\phi$ spectra will be presented.  In sect. \ref{sec:mechanism},  we will discuss the scaling behaviour of the strange particle spectra in the framework of the colour string percolation model. Finally, the conclusion is given in sect. \ref{sec:conclusion}.

\section{Method to search for the scaling behaviour} \label{sec:method}
As done in ref. \cite{pi_k_p_scaling}, we will search for the scaling behaviour of the $K_{S}^{0}$ $p_{\rm T}$ spectra with the following steps. A scaling variable, $z=p_{\rm T}/K$, and a scaled $p_{\rm T}$ spectrum, $\mathrm{\Phi}(z)=A(2\pi p_{\rm T})^{-1}d^{2}N/dp_{\rm T}dy|_{p_{\rm T}=Kz}$ will be defined first. Here $y$ is the rapidity of $K_{S}^{0}$, $(2\pi p_{\rm T})^{-1}d^{2}N/dp_{\rm T}dy$ is the invariant yield of $K_{S}^{0}$. With suitable scaling parameters $K$ and $A$ that depend on $\sqrt{s}$, the data points of the $K_{S}^{0}$ $p_{\rm T}$ spectra at 0.9, 2.76 and 7 TeV can be coalesced into one curve. In ref. \cite{pi_k_p_scaling}, $K$ and $A$ for the charged pion, kaon and proton spectra at 2.76 TeV were set to be 1. This choice was made due to reason that the $p_{\rm T}$ coverage of the spectra at 2.76 TeV is much larger than the coverage at 0.9 and 2.76 TeV. In this work, the $K_{S}^{0}$ spectrum at 2.76 TeV covers a $p_{\rm T}$ range from 0.225 to 19 GeV/c, which is larger the ranges of the spectra at 0.9 and 7 TeV, 0.1 to 9 GeV/c and  0.1 to 9 GeV/c. Therefore, to keep the similarity and consistency with ref. \cite{pi_k_p_scaling}, we prefer to set the $K$ and $A$ for the $K_{S}^{0}$ spectrum at 2.76 TeV to be 1. $K$ and $A$ values at 0.9 and 7 TeV will be determined by the quality factor method \cite{QF_1, QF_2}. Obviously, the scaling function $\mathrm{\Phi}(z)$ depends on the choice of $K$ and $A$ at 2.76 TeV. This arbitrariness could be eliminated if the spectra are presented in $u=z/\langle z \rangle=p_{\rm T}/\langle p_{\rm T} \rangle$. Here $\langle z \rangle=\int^{\infty}_{0}z\mathrm{\Phi}(z)zdz\big/\int^{\infty}_{0}\mathrm{\Phi}(z)zdz$. The  normalized scaling function then is $\mathrm{\Psi}(u)=\langle z \rangle^{2}\mathrm{\Phi}(\langle z \rangle u)\big/\int^{\infty}_{0}\mathrm{\Phi}(z)zdz$. With $\mathrm{\Psi}(u)$, the spectra at 0.9 and 7 TeV can be parameterized as $f(p_{\rm T})=\int^{\infty}_{0}\mathrm{\Phi}(z)zdz/(A\langle z \rangle^{2})\mathrm{\Psi} \left(p_{\rm T}/(K\langle z \rangle \right))$, where $K$ and $A$ are the scaling parameters at these energies. The methods to search for the scaling behaviour of the $\rm \Lambda$, $\rm \Xi$ and $\phi$ spectra are similar to that for the $K_{S}^{0}$ spectra.

\section{Scaling behaviour of the $K_{S}^{0}$, $\rm \Lambda$, $\rm \Xi$ and $\phi$ $p_{\rm T}$ spectra}
\label{sec:scaling_behaviour}
The $K_{S}^{0}$, $\rm \Lambda$ and $\rm \Xi$ $p_{\rm T}$ spectra in pp collisions at 0.9 and 7 TeV were published by the CMS  collaboration \cite{strange_production_1}. Here $\rm \Lambda$ and $\rm \Xi$ refer to $(\rm \Lambda+ \bar \Lambda)/2$ and $(\rm \Xi^{+}+\Xi^{-})/2$ respectively. For the $K_{S}^{0}$, $\rm \Lambda$ and $\rm \Xi$ $p_{\rm T}$ spectra at 2.76 TeV, so far there are no official data. As the $K_{S}^{0}$ spectrum is theoretically the same as the charged kaon spectrum, and the charged kaon spectrum at 2.76 TeV were officially published in ref. \cite{charged_kaon_spectra_2_76_TeV}, we utilize the charged kaon spectrum instead of the $K_{S}^{0}$ spectrum at this energy. For the $\rm \Lambda$ and $\rm \Xi$ spectra at 2.76 TeV, we use the preliminary results of the ALICE collaboration at this energy instead. They are publicly available in refs. \cite{strange_production_2, strange_production_3}.  The $\phi$ spectra at 0.9, 2.76 and 7 TeV were published by the ALICE collaboration \cite{strange_production_4, strange_production_5, strange_production_6}. Since the scaling parameters $K$ and $A$ at 2.76 TeV are chosen to be 1, the scaling function $\mathrm{\Phi}(z)$ is exactly the $K_{S}^{0}$, $\rm \Lambda$, $\rm \Xi$ or $\phi$ $p_{\rm T}$ spectrum at this energy. As described in ref. \cite{Tsallis_distribution_1}, due to the reason that the temperature of the hadronizing system fluctuates from event to event, the $p_{\rm T}$ spectrum of final state hadrons produced in high energy collisions follows a non-extensive statistical distribution, the Tsallis distribution \cite{Tsallis_distribution_2}. Thus, the scaling function $\mathrm{\Phi}(z)$ for strange particles can be parameterized as follows \cite{pi_k_p_scaling}
\begin{eqnarray}
\mathrm{\Phi}(z)=C_{q}\left[1-(1-q)\frac{\sqrt{m^{2}+z^{2}}-m}{z_{0}}\right]^{\frac{1}{1-q}},
\label{eq:phi_z_pt_spectrum_pp}
\end{eqnarray}
where $C_{q}$, $q$ and $z_{0}$ are free parameters, $1-q$ is a measure of the non-extensivity, $m$ is the strange particle mass. In eq. (\ref{eq:phi_z_pt_spectrum_pp}), $1/(q-1)$ determines the power law behaviour of $\mathrm{\Phi}(z)$ in the high $p_{\rm T}$ region, while $z_{0}$ controls the exponential behaviour in the low  $p_{\rm T}$ region.  $C_{q}$, $q$ and $z_{0}$ are determined by the least squares fitting of $\mathrm{\Phi}(z)$ to the $K_{S}^{0}$, $\rm \Lambda$, $\rm \Xi$ and $\phi$ $p_{\rm T}$ spectra at 2.76 TeV. The statistical and systematic errors of the data points have been added in quadrature in the fits. Table \ref{tab:id_particles_fit_parameters} tabulates $C_{q}$, $q$, $z_{0}$ and their uncertainties returned by the fits. The $\chi^2$s per degrees of freedom (dof), named reduced $\chi^{2}$s, for these fits are also given in the table.

\begin{table}[H]
\caption{\label{tab:id_particles_fit_parameters} $C_{q}$, $q$ and $z_{0}$ of $\rm{\Phi}$$(z)$ for the $K_{S}^{0}$, $\rm \Lambda$, $\rm \Xi$ and $\phi$ spectra. The uncertainties quoted are due to the quadratic sum of the statistical and systematic errors of the data points. The last column shows the reduced $\chi^{2}$s for the fits on the strange $p_{\rm T}$ spectra at 2.76 TeV. }
%\begin{center}
\begin{tabular}{@{}ccccc}
\hline\noalign{\smallskip}
\textrm{\ }&
\textrm{$C_{q}$}&
\textrm{$q$}&
\textrm{$z_{0}$ (GeV/c)}&
\textrm{$\chi^{2}$/dof}
\\
\hline
$K_{S}^{0}$ & (214$\pm$2)$\times 10^{-3}$ & 1.1402$\pm$0.0004& 0.193$\pm$0.001&12.91/55\\
$\rm \Lambda$&(21$\pm$1)$\times 10^{-3}$&1.106$\pm$0.005&0.260$\pm$0.008&6.63/26 \\
$\rm \Xi$& (169$\pm$3)$\times 10^{-5}$&1.104$\pm$0.003& 0.300$\pm$0.004&3.28/11\\
$\phi$&(96$\pm$4)$\times 10^{-4}$ &1.141$\pm$0.004&0.263$\pm$0.007&6.25/18\\
\hline
\end{tabular}
%\end{center}
\end{table}

\begin{figure}[h]
\centering
\resizebox{0.39\textwidth}{!}{\includegraphics{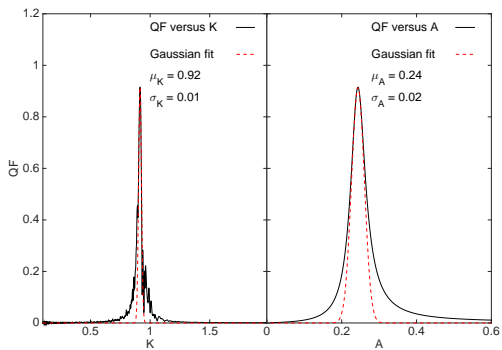}}
\caption{\label{fig:QF_vs_a_k_ks} Left (right) panel: QF versus $K$ ($A$) for $K_{S}^{0}$ at 0.9 TeV, with $A$ ($K$) fixed to 0.24 (0.92).  The black solid curve is the QF scatter plot, the red dash curve is the Gaussian fit of the peak with $\rm QF>0.91$.}
\end{figure}

As described in sect. \ref{sec:method}, the scaling parameters $K$ and $A$ at 0.9 and 7 TeV will be evaluated with the quality factor (QF) method. Compared with the method utilized in ref. \cite{inclusive_scaling}, this method is more robust since it does not rely on the shape of the scaling function. To define the quality factor, a set of data points ($\rho^{i}, \tau^{i}$) is considered first. Here $\rho^{i}=p_{\rm T}^{i}/K$, $\tau^{i}=\textrm{log}(A(2\pi p^{i}_{\rm T})^{-1}d^{2}N^{i}/dp^{i}_{\rm T}dy^{i})$, $\rho^{i}$ are ordered, $\tau^{i}$ are rescaled so that they are in the range between 0 and 1. Then, the QF is introduced as follows \cite{QF_1,QF_2}
\begin{eqnarray}
\textrm{QF}(K,A)=\left[\sum_{i=2}^{n}\frac{(\tau^{i}-\tau^{i-1})^{2}}{(\rho^{i}-\rho^{i-1})^{2}+1/n^{2}}\right]^{-1},
\label{eq:QF_definition}
\end{eqnarray}
where $n$ is the number of data points and $1/n^{2}$ keeps the sum finite in the case of  two points taking the same $\rho$ value. It is obvious that a large contribution to the sum in the QF is given if two successive data points are close in $\rho$ and far in $\tau$. Therefore, a set of data points are expected to lie close to a single curve if they have a small sum (a large QF) in eq. (\ref{eq:QF_definition}). The best set of ($K$,  $A$) at 0.9 (7) TeV is chosen to be the one which globally maximizes the QF of the data points at 0.9 (7) and 2.76 TeV. Table \ref{tab:id_particles_a_k_parameters} tabulates $K$ and $A$ for the $K_{S}^{0}$, $\rm \Lambda$, $\rm \Xi$ and $\phi$ spectra at 0.9, 2.76 and 7 TeV. Also shown in the table is the maximum QF ($\rm QF_{max}$). In order to determine the uncertainties of $K$ and $A$ at 0.9 and 7 TeV, we utilize the method mentioned in ref. \cite{QF_1}. Let's take the determination of the uncertainty of $K$ ($A$) for $K_{S}^{0}$ at 0.9 TeV as an example. In fig. \ref{fig:QF_vs_a_k_ks} we first plot the QF as a function of $K$ ($A$) with $A$ ($K$) fixed to the value 0.24 (0.92) returned by the QF method. The peak value with $\rm QF >(QF_{max}-0.01)$ shows a good scaling and we make a Gaussian fit to this bump. The standard deviation of the Gaussian fit, $\sigma_{K(A)}$, is taken as the uncertainty of $K$ ($A$) for $K_{S}^{0}$ at 0.9 TeV. The mean value of the Gaussian fit, $\mu_{K(A)}$, is consistent with the value of $K$ ($A$) returned by the QF method, thus this method to determine the uncertainties of scaling parameters is robust. The errors of $K$ and $A$ for $K_{S}^{0}$ at 7 TeV, $\rm \Lambda$, $\rm \Xi$ and $\phi$ at 0.9 and 7 TeV are determined by making  Gaussian fits to the peaks with $\rm QF>(QF_{max}-0.01)$.

\begin{table}[h]
\caption{\label{tab:id_particles_a_k_parameters} $K$ and $A$ for the $K_{S}^{0}$, $\rm \Lambda$, $\rm \Xi$ and $\phi$ spectra at 0.9, 2.76 and 7 TeV. The $\rm QF_{max}$ is shown in the last column of the table. The standard deviations of the Gaussian fits to the peaks of the QF scatter plots at 0.9 and 7 TeV are taken as the uncertainties of $K$ and $A$ at these two energies.}
\begin{center}
\begin{tabular}{@{}cccccc}
\hline
 \textrm{\ }&
\textrm{$\sqrt{s}$ (TeV)}&
\textrm{$K$}&
\textrm{$A$}&
$\rm QF_{max}$\\
\hline
\textrm{\ }& 0.9 &0.92$\pm$0.01& 0.24$\pm$0.02&0.92\\
\textrm{$K_{S}^{0}$}& 2.76 &1 & 1&-\\
\textrm{\ }& 7 & 1.14$\pm$0.01 & 0.23$\pm$0.02&0.92\\
\hline
\textrm{\ }& 0.9 & 0.85$\pm$0.01 & 0.20$\pm$0.02&1.46\\
\textrm{$\rm \Lambda$}& 2.76& 1&1&-\\
\textrm{\ }& 7 & 1.10$\pm$0.02 & 0.20$\pm$0.02&1.21\\
\hline
\textrm{\ }& 0.9 & 0.86$\pm$0.02 &0.20$\pm$0.03&1.94\\
\textrm{$\rm \Xi$}& 2.76 &1 & 1&-\\
\textrm{\ }& 7 & 1.13$\pm$0.02 & 0.20$\pm$0.02&2.16\\
\hline
\textrm{\ }& 0.9 & 0.85$\pm$0.06 &0.89$\pm$0.17&6.96\\
\textrm{$\phi$}& 2.76 &1 & 1&-\\
\textrm{\ }& 7 & 1.06$\pm$0.03 & 0.91$\pm$0.07&2.89\\
\hline
\end{tabular}
\end{center}
\end{table}

\begin{figure}[h]
\centering
\resizebox{0.39\textwidth}{!}{\includegraphics{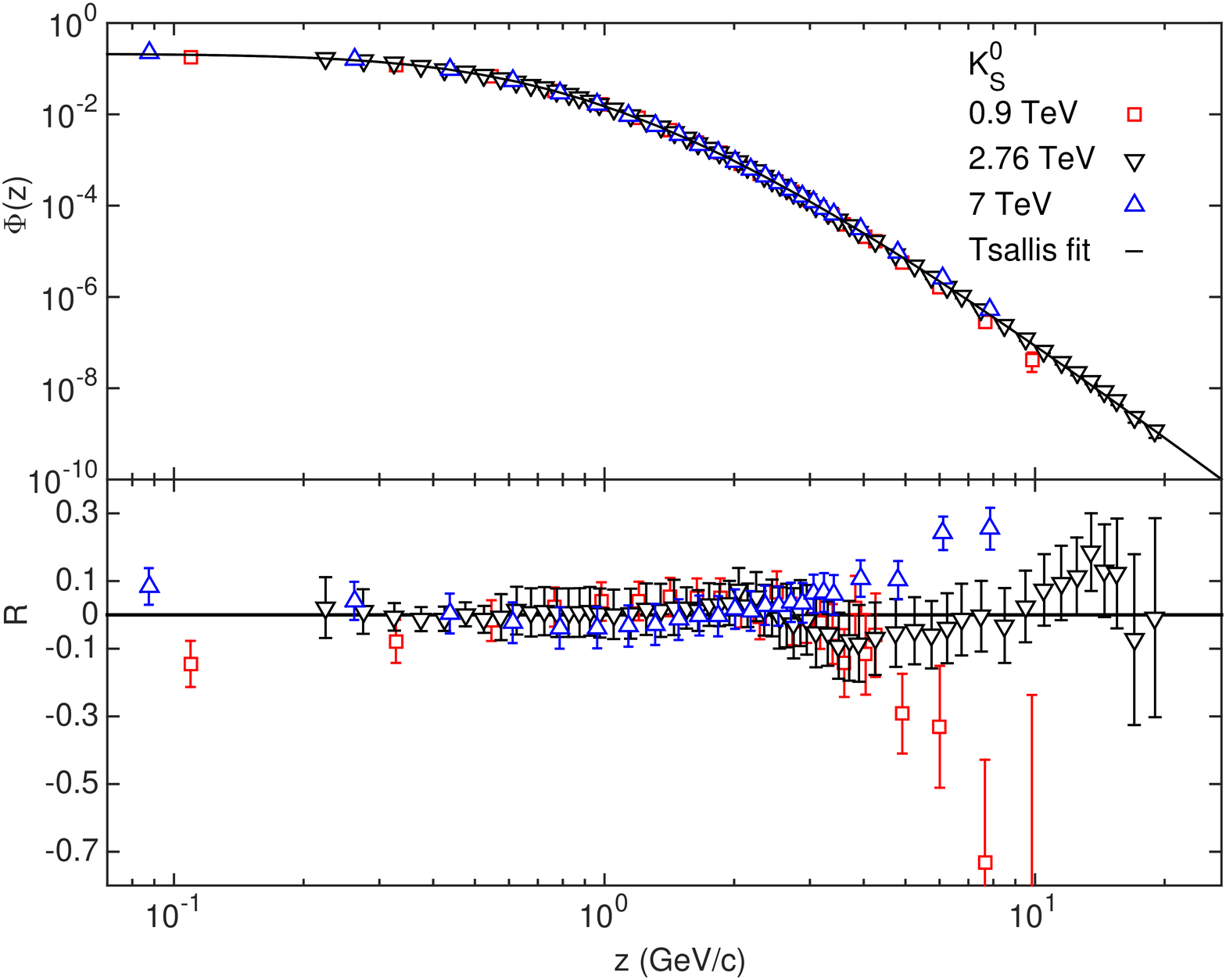}}
\caption{\label{fig:ks_z_plus_ratio_log_2_76_TeV_QF}Upper panel: the scaling behaviour of the $K_{S}^{0}$  $p_{\rm T}$ spectra presented in $z$ at 0.9, 2.76 and 7 TeV. The solid curve is from $\mathrm{\Phi}(z)$ with parameters in the second row of table \ref{tab:id_particles_fit_parameters}. The data points are taken from refs. \cite{strange_production_1, charged_kaon_spectra_2_76_TeV}. Lower panel: the $R$ distributions. The $R$ value for the last data point at 0.9 TeV is -1.27 and it is not shown in the lower panel.}
\end{figure}

Using the scaling parameters $K$ and $A$ in table \ref{tab:id_particles_a_k_parameters}, now we can shift the $K_{S}^{0}$ $p_{\rm T}$ spectra at 0.9 and 7 TeV to the spectrum at 2.76 TeV. They are shown in the upper panel of fig. \ref{fig:ks_z_plus_ratio_log_2_76_TeV_QF}. On a log scale, most of the data points at different energies appear consistent with the universal curve which is described by $\mathrm{\Phi}(z)$ in eq. (\ref{eq:phi_z_pt_spectrum_pp}) with parameters in the second row of table \ref{tab:id_particles_fit_parameters}. In order to see how well the data points agree with the fitted curve, a ratio, $R=\rm (data-fitted)/data$, is evaluated at 0.9, 2.76 and 7 TeV. The uncertainty of $R$ is determined to be $\rm (fitted/data)\times(\Delta data/data)$, where $\rm \Delta data$ is the total uncertainty of the data point. The $R$ distribution is shown in the lower panel of the figure. Except for the last three points in the high $p_{\rm T}$ region at 0.9 TeV, all the other points have $R$ values in the range between -0.3 and 0.3, which implies that the agreement between the data points and the fitted curve is within 30$\%$. This agreement roughly corresponds to the systematic errors on $R$ and the accuracy of the fits. If we take into account the systematic errors on $R$, then this agreement is within 22$\%$.

\begin{figure}[h]
\centering
\resizebox{0.39\textwidth}{!}{\includegraphics{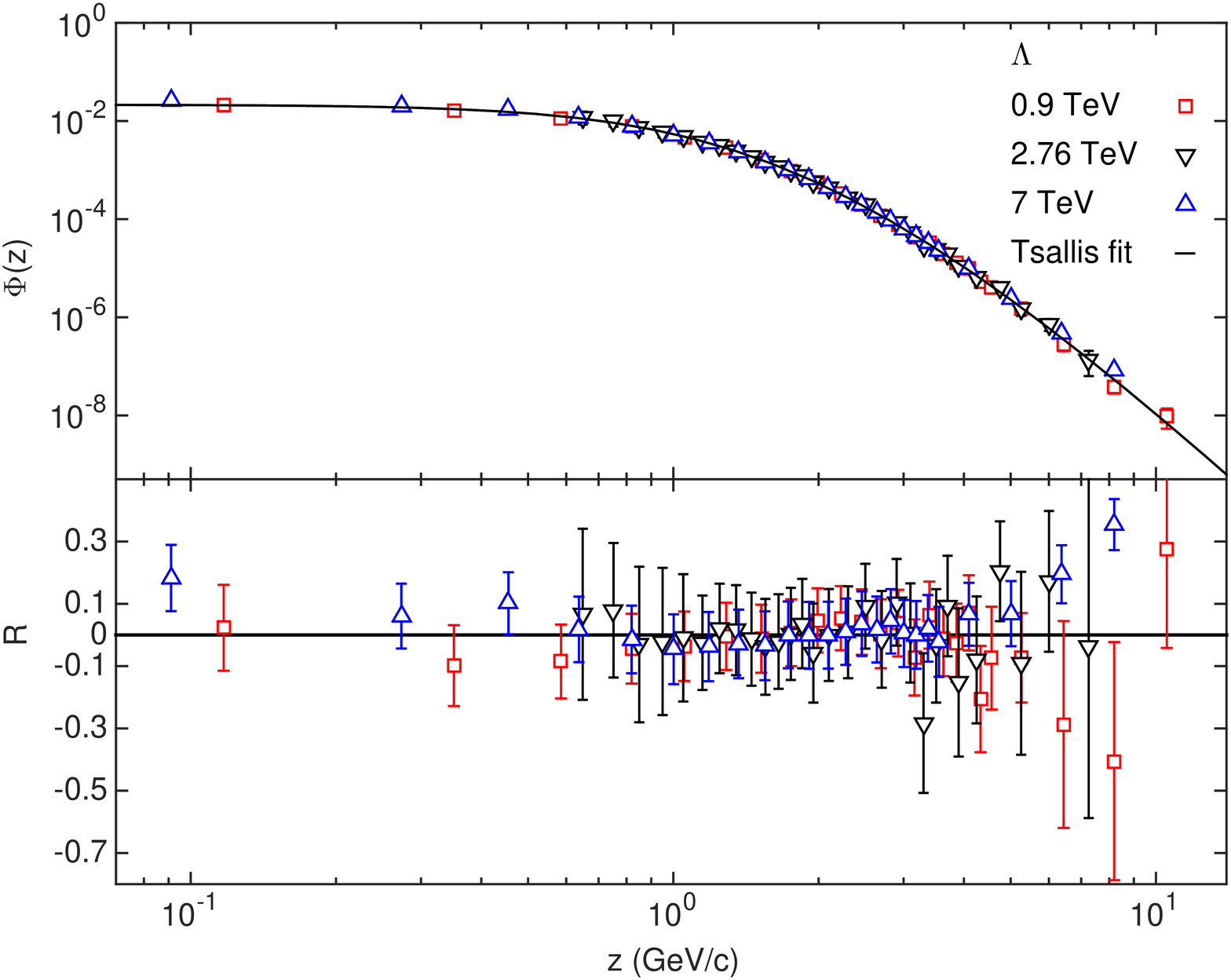}}
\caption{\label{fig:lambda_z_plus_ratio_log_2_76_TeV_QF}Upper panel: the scaling behaviour of the $\rm \Lambda$ $p_{\rm T}$ spectra presented in $z$ at  0.9, 2.76 and 7 TeV. The solid curve is from $\mathrm{\Phi}(z)$ with parameters in the third row of table \ref{tab:id_particles_fit_parameters}. The data points are taken from refs. \cite{strange_production_1, strange_production_2}. Lower panel: the $R$ distributions.}
\end{figure}

\begin{figure}[h]
\centering
\resizebox{0.39\textwidth}{!}{\includegraphics{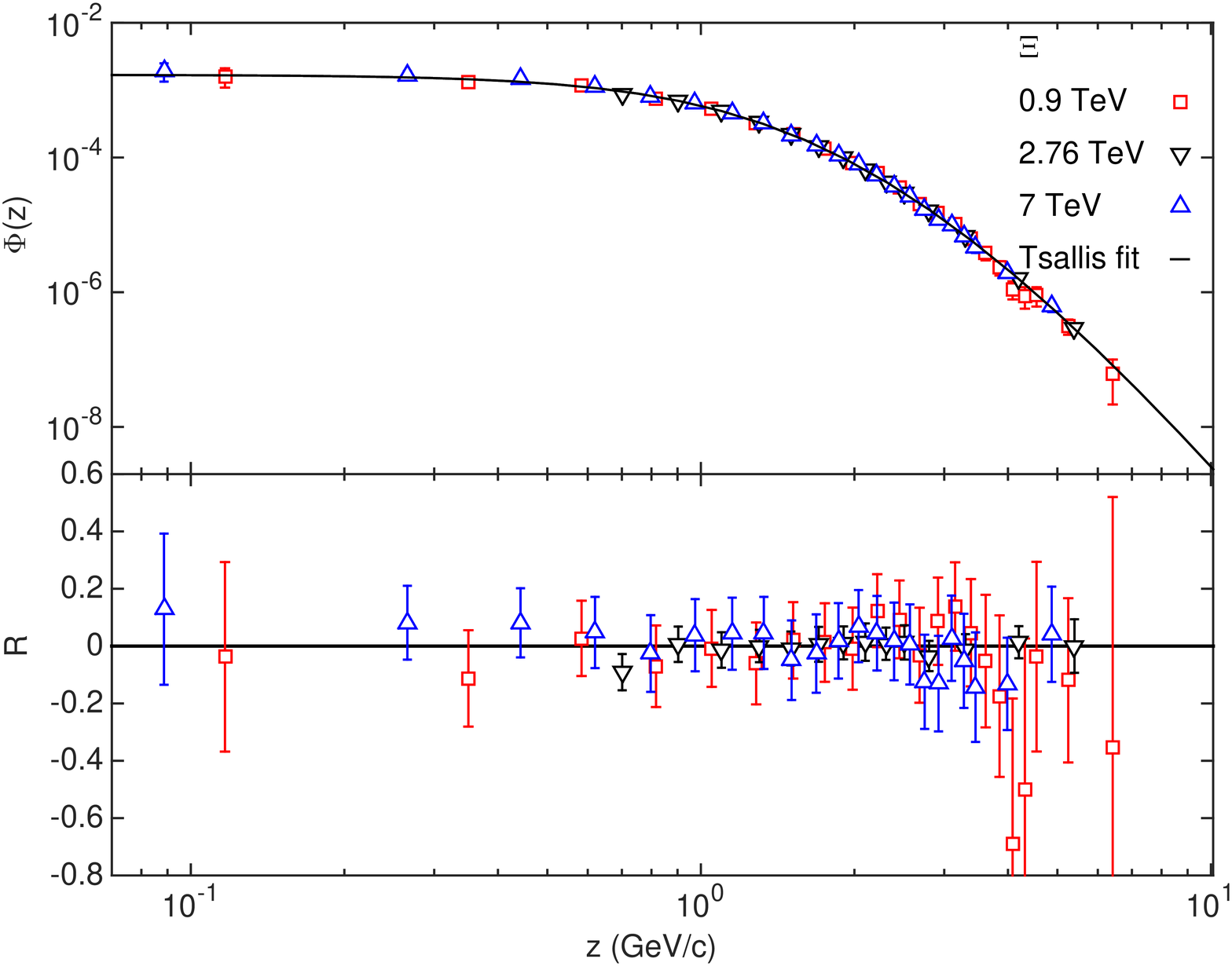}}
\caption{\label{fig:xi_z_plus_ratio_log_2_76_TeV_QF}Upper panel: the scaling behaviour of the $\rm \Xi$ $p_{\rm T}$ spectra presented in $z$ at  0.9, 2.76 and 7 TeV. The solid curve is from $\mathrm{\Phi}(z)$ with parameters in the fourth row of table \ref{tab:id_particles_fit_parameters}. The data points are taken from refs. \cite{strange_production_1,  strange_production_3}. Lower panel: the $R$ distributions. }
\end{figure}

\begin{figure}[h]
\centering
\resizebox{0.39\textwidth}{!}{\includegraphics{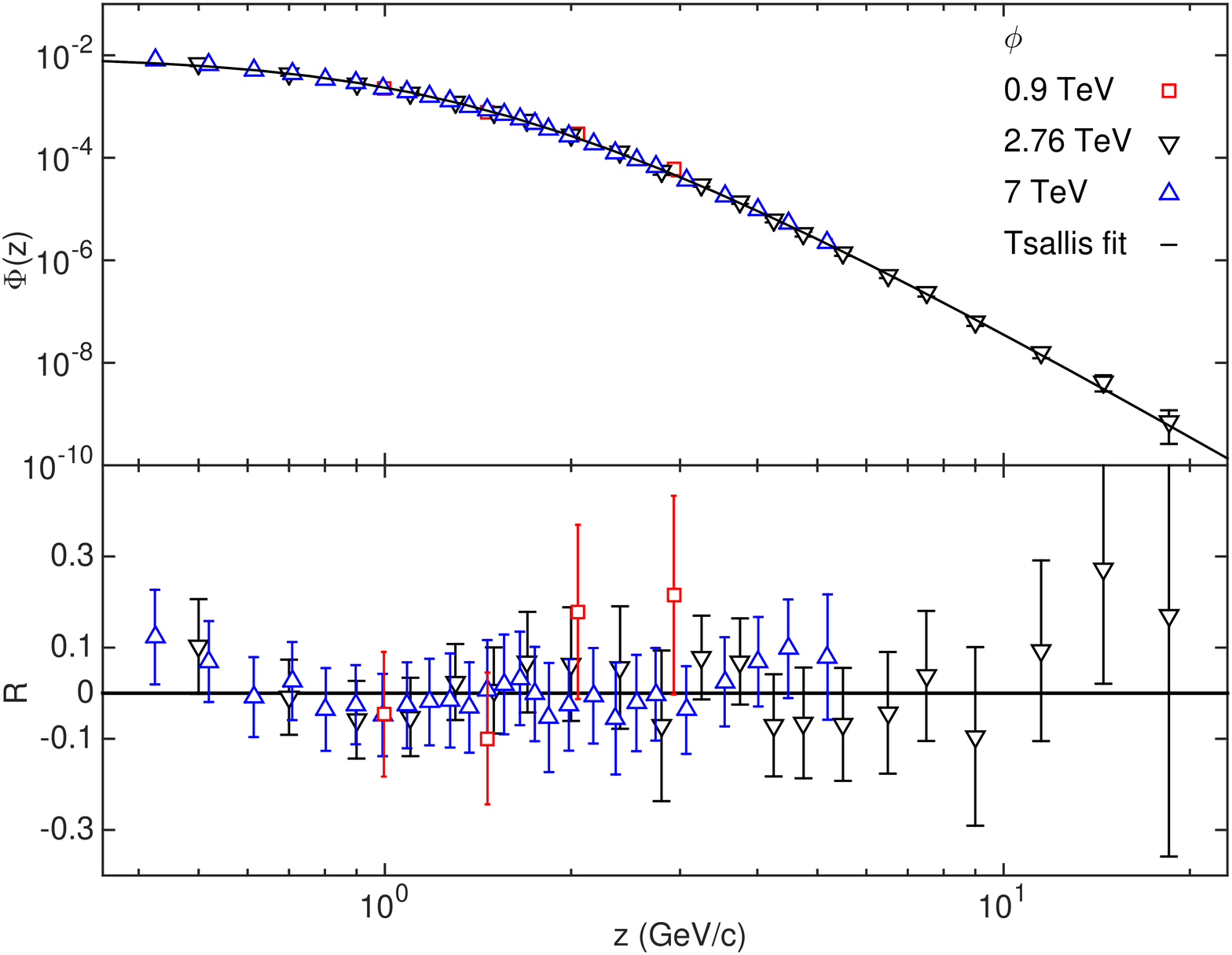}}
\caption{\label{fig:phi_z_plus_ratio_log_2_76_TeV_QF}Upper panel: the scaling behaviour of the $\phi$ $p_{\rm T}$ spectra presented in $z$ at  0.9, 2.76 and 7 TeV. The solid curve is from $\mathrm{\Phi}(z)$ with parameters in the fifth row of table \ref{tab:id_particles_fit_parameters}. The data points are taken from refs. \cite{strange_production_4, strange_production_5, strange_production_6}. Lower panel: the $R$ distributions. }
\end{figure}

In the upper panels of figs. \ref{fig:lambda_z_plus_ratio_log_2_76_TeV_QF}, \ref{fig:xi_z_plus_ratio_log_2_76_TeV_QF} and \ref{fig:phi_z_plus_ratio_log_2_76_TeV_QF}, we present the scaling behaviour of the $\rm \Lambda$, $\rm \Xi$ and $\phi$ $p_{\rm T}$ spectra at 0.9, 2.76 and 7 TeV. In the lower panels of these figures are the $R$ distributions for these spectra. For the $\rm \Lambda$ spectra, except for the second-to-last point  at 0.9 TeV and the last point at 7 TeV, all the other points agree with the fitted curve within 30$\%$. Taking into account the systematic uncertainties of $R$, this agreement is within 11$\%$. For the $\rm \Xi$ spectra, except for the points with $z=$ 4.1, 4.3 and 6.4 GeV/c at 0.9 TeV, all the other points are consistent with the fitted curve within 20$\%$. Taking into account the systematic errors of $R$, this consistency is within 3$\%$. For the $\phi$ spectra, all the points are in agreement with the fitted curve within 30$\%$. With the consideration of the systematic errors of $R$, this agreement is within 18$\%$.

\begin{figure}[h]
\centering
\resizebox{0.39\textwidth}{!}{\includegraphics{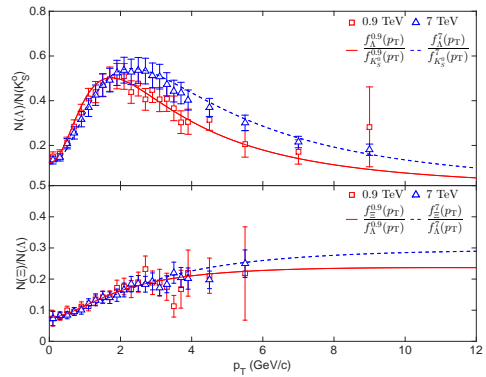}}
\caption{\label{fig:ratio_xi_lambda_and_lambda_ks}Upper (Lower) panel: the $N(\mathrm{\Lambda})/N(K_{S}^{0})$ ($N(\mathrm{\Xi})/N(\mathrm{\Lambda})$) distributions at 0.9 (solid line) and 7 TeV (dashed line). The data points are taken from ref. \cite{strange_production_1}. The values of $N(\mathrm{\Lambda})/N(K_{S}^{0})$ for data points have been divided by 2, since here $\rm \Lambda$ refers to $\rm (\Lambda+\bar \Lambda)/2$.}
\end{figure}

From the above statement, we have shown that the $p_{\rm T}$ spectra of $K_{S}^{0}$, $\rm \Lambda$, $\rm \Xi$ and $\phi$ at 0.9, 2.76 and 7 TeV exhibit a scaling behaviour independent of $\sqrt{s}$. As described in sect. \ref{sec:method}, the scaling function $\mathrm{\Phi}(z)$ relies on $K$ and $A$ chosen at 2.76 TeV. In order to get rid of this reliance, we utilize the scaling variable $u=z/\langle z \rangle$ instead. The $\langle z \rangle$ values for the $K_{S}^{0}$, $\rm \Lambda$, $\rm \Xi$ and $\phi$ $p_{\rm T}$ spectra are determined  as 0.701$\pm$0.008, 0.97$\pm$0.03, 1.12$\pm$0.01 and 1.04$\pm$0.02 GeV/c, where the errors are due to the uncertainties of $C_{q}$, $q$ and $z_{0}$ in table \ref{tab:id_particles_fit_parameters}. The corresponding normalized scaling function $\mathrm{\Psi}(u)$ is
\begin{eqnarray}
\mathrm{\Psi}(u)=C^{\prime}_{q}\left[1-(1-q^{\prime})\frac{\sqrt{(m^{\prime})^{2}+u^{2}}-m^{\prime}}{u_{0}}\right]^{\frac{1}{1-q^{\prime}}}.
\label{eq:psi_u_pt_spectrum_pp}
\end{eqnarray}
Here $C^{\prime}_{q}=\langle z \rangle^{2}C_{q}/\int^{\infty}_{0}\Phi(z)zdz$, $q^{\prime}=q$, $u_{0}=z_{0}/\langle z \rangle$ and $m^{\prime}=m/\langle z \rangle$. Their values  are presented in table \ref{tab:id_particles_normalized_parameters}. As described in sect. \ref{sec:method}, with $\mathrm{\Psi}(u)$, the spectra of $K_{S}^{0}$, $\rm \Lambda$, $\rm \Xi$ and $\phi$ at 0.9 (7) TeV can be parameterized as $f(p_{\rm T})=\int^{\infty}_{0}\mathrm{\Phi}(z)zdz/(A\langle z \rangle^{2})\mathrm{\Psi} \left(p_{\rm T}/(K\langle z \rangle \right))$, where $K$ and $A$ are the scaling parameters of these strange particles at 0.9 (7) TeV in table \ref{tab:id_particles_a_k_parameters}. In ref. \cite{strange_production_1}, the CMS collaboration have presented the relative production versus $p_{\rm T}$ between different strange particle species, $N(\mathrm{\Lambda})/N(K_{S}^{0})$ and $N(\mathrm{\Xi})/N(\mathrm{\Lambda})$, at 0.9 and 7 TeV. In the upper (lower) panel of fig. \ref{fig:ratio_xi_lambda_and_lambda_ks}, we show that the $N(\mathrm{\Lambda})/N(K_{S}^{0})$ ($N(\mathrm{\Xi})/N(\mathrm{\Lambda})$) distributions in data at 0.9 and 7 TeV are well described by $f_{\mathrm{\Lambda}}(p_{\rm T})/f_{K_{S}^{0}}(p_{\rm T})$  ($f_{\mathrm{\Xi}}(p_{\rm T})/f_{\mathrm{\Lambda}}(p_{\rm T})$). This agreement is a definite indication that the scaling behaviour exists in the $p_{\rm T}$ spectra of strange particles at 0.9, 2.76 and 7 TeV.  The $p_{\rm T}$ dependence of the relative production can be explained as follows. At  low $p_{\rm T}$, $f(p_{\rm T})$ inclines to be an exponential distribution which is controlled by the parameter $z_{0}=u_{0}\langle z \rangle$. For $N(\mathrm{\Lambda})/N(K_{S}^{0})$ ($N(\mathrm{\Xi})/N(\mathrm{\Lambda}))$, the $z_{0}$ value for $\rm \Lambda$ ($\rm \Xi$) is larger than that for $K_{S}^{0}$ ($\rm \Lambda$), therefore both $N(\mathrm{\Lambda})/N(K_{S}^{0})$ and $N(\mathrm{\Xi})/N(\mathrm{\Lambda})$ grow with  $p_{\rm T}$. At high $p_{\rm T}$, $f(p_{\rm T})$ prefers to be a power law distribution which is dominated by $1/(q^{\prime}-1)$. $q^{\prime}$ value for $\rm \Lambda$ ($\rm \Xi$) is smaller than (almost equal to) that for $K_{S}^{0}$ ($\rm \Lambda$), thus $N(\mathrm{\Lambda})/N(K_{S}^{0})$ decreases with $p_{\rm T}$ while $N(\mathrm{\Xi})/N(\mathrm{\Lambda})$ appears to be flat.

\begin{table}[H]
  \caption{\label{tab:id_particles_normalized_parameters} $C^{\prime}_{q}$, $q^{\prime}$, $u_{0}$ and $m^{\prime}$ of $\mathrm{\Psi}(u)$ for $K_{S}^{0}$, $\rm \Lambda$, $\rm \Xi$ and $\phi$. The uncertainties quoted are due to the errors of $C_{q}$, $q$ and $z_{0}$ in table \ref{tab:id_particles_fit_parameters}. }
%\begin{center}
\begin{tabular}{@{}ccccc}
\hline\noalign{\smallskip}
\textrm{\ }&
$C^{\prime}_{q}$&
$q^{\prime}$&
$u_{0}$&
$m^{\prime}$
\\
\hline
${K}_{S}^{0}$ &2.86$\pm$0.03&1.1402$\pm$0.0004&0.275$\pm$0.004&0.704$\pm$0.008\\
$\rm \Lambda$&2.23$\pm$0.04&1.106$\pm$0.005&0.268$\pm$0.004&1.15$\pm$0.03\\
$\rm \Xi$&2.21$\pm$0.02&1.104$\pm$0.003&0.267$\pm$0.002&1.32$\pm$0.01\\
$\phi$&2.54$\pm$0.04&1.141$\pm$0.004&0.254$\pm$0.003&0.98$\pm$0.02\\
\hline
\end{tabular}
%\end{center}
\end{table}

\section{Discussions}
\label{sec:mechanism}
%The mean transverse momentum squared $\langle p_{\rm T}^{2}\rangle_{n}$ of strange particles produced by a cluster with $n$ strings is given by $\langle p_{\rm T}^{2}\rangle_{n}=\sqrt{nS_{1}/S_{n}} \langle p_{\rm T}^{2}\rangle_{1}$, where $\langle p_{\rm T}^{2}\rangle_{1}$ is the mean $p_{\rm T}^{2}$ of strange particles produced by a single string, $S_{n}$ is the transverse area of the cluster and $nS_{1}/S_{n}$ is the degree of string overlap. 
In sect. \ref{sec:scaling_behaviour}, we have shown that there is indeed a scaling behaviour in the $K_{S}^{0}$, $\rm \Lambda$, $\rm \Xi$ and $\phi$ $p_{\rm T}$ spectra in pp collisions at 0.9, 2.76 and 7 TeV. This scaling behaviour appears when the spectra are presented in terms of the scaling variable $z$. Now we would like to discuss this scaling behaviour in terms of the colour string percolation (CSP) model \cite{string_perco_model_1, string_perco_model_2}.

In this model, colour strings are stretched between the partons of the projectile and target protons in pp collisions. These strings then will split into new ones by the production of sea $q\bar{q}$ pairs from the vacuum. Strange particles such as $K_{S}^{0}$, $\rm \Lambda$, $\rm \Xi$ and $\phi$ are produced through the hadronization of these new strings. In the transverse plane, the colour strings look like discs, each of which has an area, $S_{1}=\pi r_{0}^{2}$, $r_{0}\approx 0.2$ fm. When the collision energy increases, the number of strings grows and they interact with each other and start to overlap to form clusters. A cluster with $n$ strings is assumed to behave as a single string. The colour field of the cluster $\vv{Q}_{n}$ is the vectorial sum of the colour charge of each individual $\vv{Q}_{1}$ string, $\vv{Q}_{n}=\sum_{1}^{n}\vv{Q}_{1}$. Since the individual string colour fields are oriented arbitrarily, the average value of $\vv{Q}_{1i}\cdot\vv{Q}_{1j}$ is zero and $\vv{Q}_{n}^{2}=n\vv{Q}_{1}^{2}$. $\vv{Q}_{n}$ also depends on the transverse area of each individual string $S_{1}$ and the transverse area of the cluster $S_{n}$. Thus, $Q_{n}=\sqrt{nS_{n}/S_{1}} Q_{1}$. As the multiplicity of strange particles produced from the cluster is proportional to its colour charge, $\mu_{n}=\sqrt{nS_{n}/S_{1}} \mu_{1}$, where $\mu_{1}$ is the multiplicity of strange particles produced by a single string. Since the transverse momentum is conserved before and after the overlapping, $\mu_{n}\langle p_{\rm T}^{2}\rangle_{n}=n\mu_{1}\langle p_{\rm T}^{2}\rangle_{1}$, where $\langle p_{\rm T}^{2}\rangle_{n}$ is the mean $p_{\rm T}^{2}$ of strange particles produced by the cluster, $\langle p_{\rm T}^{2}\rangle_{1}$ is the mean $p_{\rm T}^{2}$ of strange particles produced by a single string. Therefore, $\langle p_{\rm T}^{2}\rangle_{n}=\sqrt{nS_{1}/S_{n}} \langle p_{\rm T}^{2}\rangle_{1}$, where $nS_{1}/S_{n}$ is the degree of string overlap. For the case where strings just get in touch with each other, $S_{n}=nS_{1}$, $nS_{1}/S_{n}=1$ and $\langle p_{\rm T}^{2}\rangle_{n}=\langle p_{\rm T}^{2}\rangle_{1}$, which means that the $n$ strings fragment into strange hadrons independently. For the case in which strings maximally overlap with each other, $S_{n}=S_{1}$, $nS_{1}/S_{n}=n$ and $\langle p_{\rm T}^{2}\rangle_{n}=\sqrt{n} \langle p_{\rm T}^{2}\rangle_{1}$, which means that the mean $p_{\rm T}^{2}$ is maximally enhanced due to the percolation. The $p_{\rm T}$ spectra of strange particles produced in pp collisions can be written as a superposition of the $p_{\rm T}$ distribution produced by each cluster, $g(x, p_{\rm T})$, weighted with the cluster's size distribution $W(x)$,
\begin{eqnarray}
\frac{d^{2}N}{2\pi p_{\rm T}dp_{\rm T}dy}=C\int_{0}^{\infty}W(x)g(x,p_{\rm T})dx,
\label{eq:CPS_formula}
\end{eqnarray}
where $C$ is a normalization parameter which characterizes the total number of clusters formed for strange particles before hadronization. $W(x)$ is supposed to be a gamma distribution,
\begin{eqnarray}
W(x)=\frac{\gamma}{\Gamma(\kappa)}(\gamma x)^{\kappa-1}\textrm{exp}(-\gamma x),
\label{eq:gamma_function}
\end{eqnarray}
where $x$ is proportional to $1/\langle p_{\rm T}^{2}\rangle_{n}$, $\kappa$ and $\gamma$ are free parameters. $\kappa$ is related to the dispersion of the size distribution, $1/\kappa=(\langle x^{2}\rangle-\langle x\rangle^{2})/\langle x\rangle^{2}$. It depends on the density of the strings, $\eta=(r_{0}/R)^{2}N_{s}$, where $R$ is the effective radius of the interaction region, $N_{s}$ is the average number of strings of the cluster. $\gamma$ is related to the mean $x$, $\langle x\rangle=\kappa/\gamma$.

\begin{figure}[h]
\centering
\resizebox{0.39\textwidth}{!}{\includegraphics{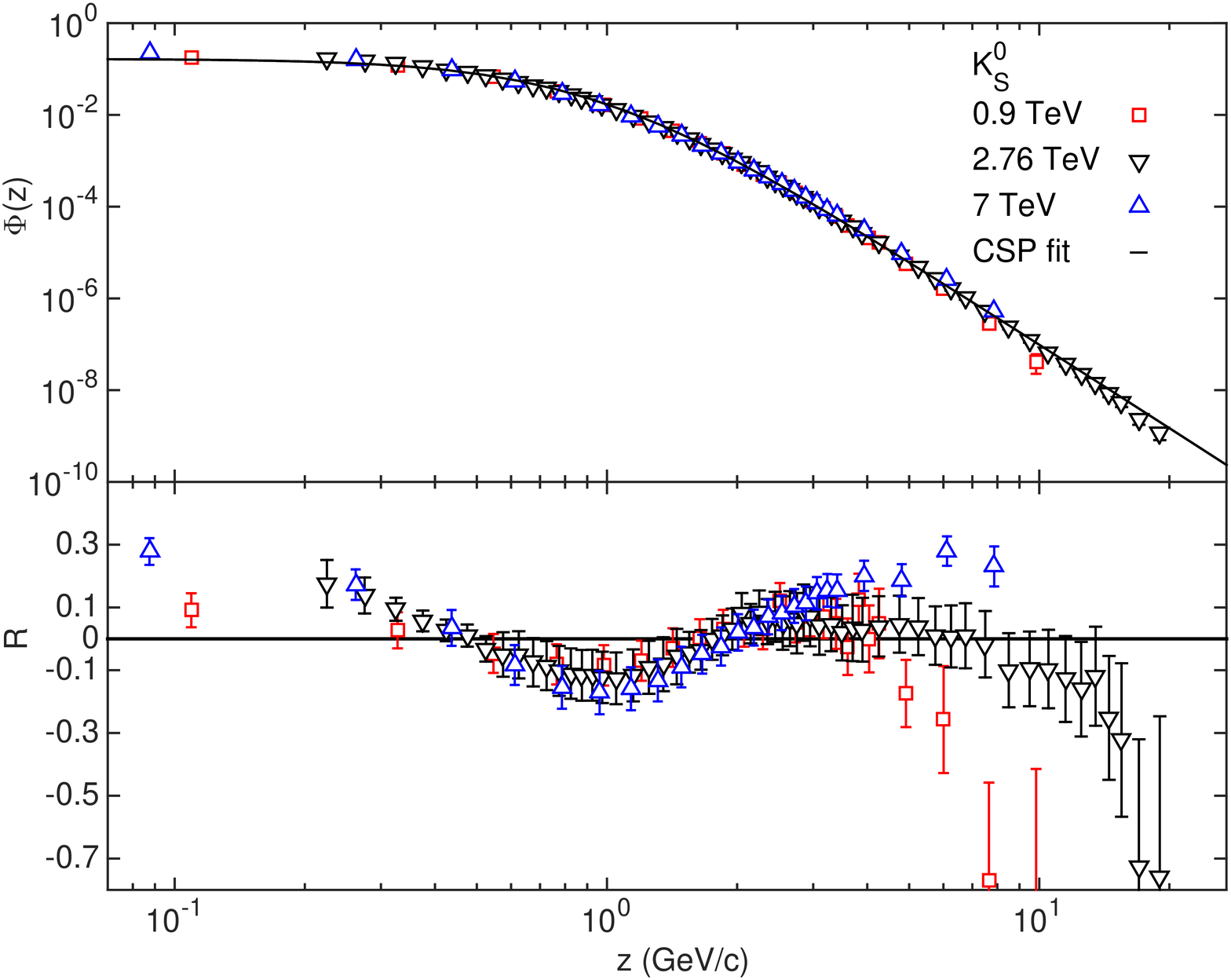}}
\caption{\label{fig:ks_csp_fit_plus_ratio_log_2_76_TeV} Upper panel: the scaling behaviour of the $K_{S}^{0}$ $p_{\rm T}$ spectra presented in $z$ at  0.9, 2.76 and 7 TeV. The solid curve is the CSP fit in eq. (\ref{eq:CPS_formula}) with parameters in the second row of table \ref{tab:CSP_comb_fit_parameters}. The data points are taken from refs. \cite{strange_production_1, charged_kaon_spectra_2_76_TeV}. Lower panel: the $R$ distributions. The $R$ value for the last data point at 0.9 TeV is -1.60 and it is not shown in the lower panel.}
\end{figure}

\begin{figure}[h]
\centering
\resizebox{0.39\textwidth}{!}{\includegraphics{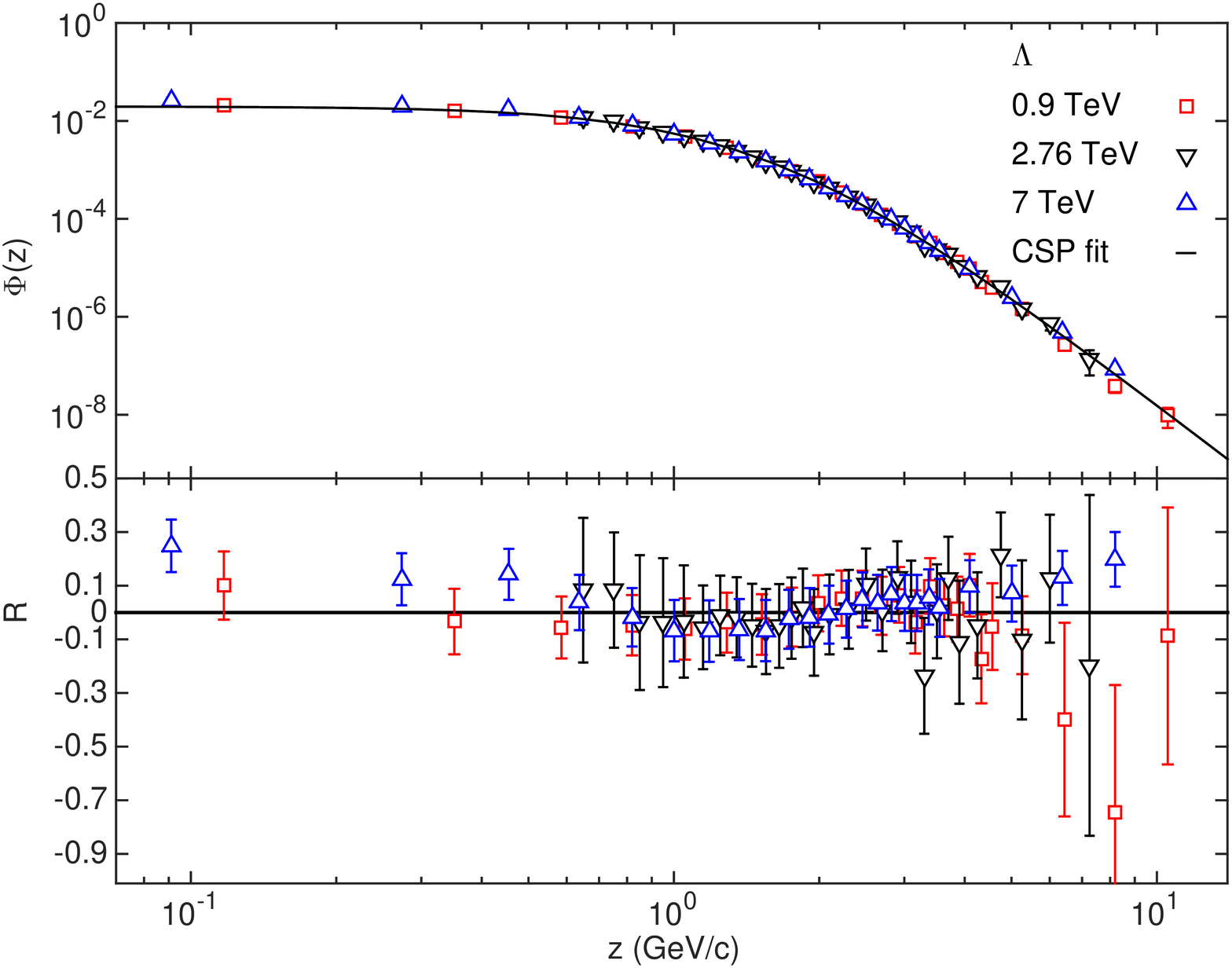}}
\caption{\label{fig:lambda_csp_fit_plus_ratio_log_2_76_TeV} Upper panel: the scaling behaviour of the $\rm \Lambda$ $p_{\rm T}$ spectra presented in $z$ at  0.9, 2.76 and 7 TeV. The solid curve is the CSP fit in eq. (\ref{eq:CPS_formula}) with parameters in the third row of table \ref{tab:CSP_comb_fit_parameters}. The data points are taken from refs. \cite{strange_production_1, strange_production_2}. Lower panel: the $R$ distributions. }
\end{figure}

\begin{figure}[h]
\centering
\resizebox{0.39\textwidth}{!}{\includegraphics{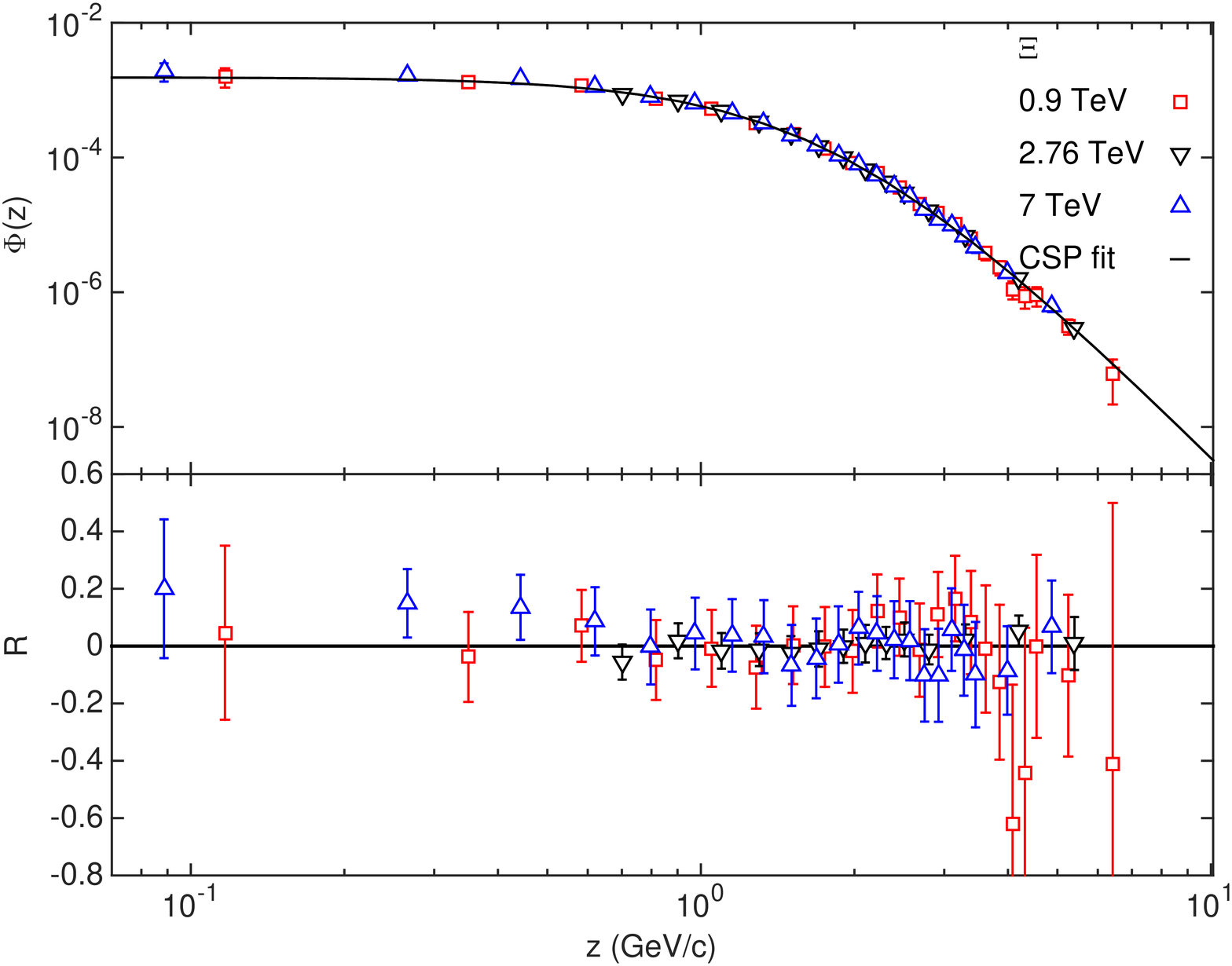}}
\caption{\label{fig:xi_csp_fit_plus_ratio_log_2_76_TeV} Upper panel: the scaling behaviour of the $\rm \Xi$ $p_{\rm T}$ spectra presented in $z$ at  0.9, 2.76 and 7 TeV. The solid curve is the CSP fit in eq. (\ref{eq:CPS_formula}) with parameters in the fourth row of table \ref{tab:CSP_comb_fit_parameters}. The data points are taken from refs. \cite{strange_production_1,  strange_production_3}. Lower panel: the $R$ distributions.}
\end{figure}

\begin{figure}[h]
\centering
\resizebox{0.39\textwidth}{!}{\includegraphics{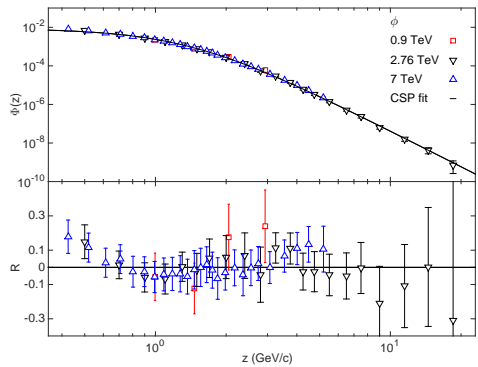}}
\caption{\label{fig:phi_csp_plus_ratio_log_2_76_TeV}Upper panel: the scaling behaviour of the $\phi$ $p_{\rm T}$ spectra presented in $z$ at  0.9, 2.76 and 7 TeV. The solid curve is the CSP fit in eq. (\ref{eq:CPS_formula}) with parameters in the fifth row of table \ref{tab:CSP_comb_fit_parameters}. The data points are taken from refs. \cite{strange_production_4, strange_production_5, strange_production_6}. Lower panel: the $R$ distributions. }
\end{figure}

In order to see whether the CSP model can describe the scaling behaviour of strange particle $p_{\rm T}$ spectra, we attempt to fit eq. (\ref{eq:CPS_formula}) to the combination of the scaled data points at 0.9, 2.76 and 7 TeV with the least squares method. Here the cluster's fragmentation function in the CSP fit is chosen as the Schwinger formula \cite{schwinger_formula}
\begin{eqnarray}
g(x, p_{\rm T})=\textrm {exp}(-p_{\rm T}^{2}x).
\label{eq:frag_function}
\end{eqnarray}
$C$, $\gamma$ and $\kappa$ returned by the fits are listed in table \ref{tab:CSP_comb_fit_parameters}. From the table, we see that the dispersion of the cluster's size distribution (1/$\kappa$) for strange mesons ($K_{S}^{0}$ and $\phi$) is larger than that of strange baryons ($\rm \Lambda$ and $\rm \Xi$), while the dispersion of the cluster's size distribution for $K_{S}^{0}$ ($\rm \Lambda$) is almost equal to that of $\phi$ ($\rm \Xi$) when considering the errors. This implies that the strange mesons and baryons are produced from clusters with different size distributions, while the strange mesons (baryons) $K_{S}^{0}$ and $\phi$ ($\rm \Lambda$ and $\rm \Xi$) originate from clusters with the same size distributions. The cluster's size distributions for strange mesons are more dispersed than those for strange baryons. The difference between the cluster's size distributions of strange mesons and baryons could be explained as follows. As described in ref. \cite{string_perco_model_1}, since additional quarks required to form a baryon are provided by the quarks of the overlapping strings that form the cluster, the baryons probe a higher string density than mesons for the same energy of collisions. When $\eta$ is above the critical string density at which the string percolation appears, $\kappa$ increases with $\eta$ \cite{string_perco_model_3}.  Therefore the $\kappa$ values for strange baryons are larger than those for strange mesons. The fit results for $K_{S}^{0}$, $\rm \Lambda$, $\rm \Xi$ and $\phi$ are presented in the upper panels of figs. \ref{fig:ks_csp_fit_plus_ratio_log_2_76_TeV}, \ref{fig:lambda_csp_fit_plus_ratio_log_2_76_TeV}, \ref{fig:xi_csp_fit_plus_ratio_log_2_76_TeV} and \ref{fig:phi_csp_plus_ratio_log_2_76_TeV} respectively. The $R$ distributions are shown in the lower panels of these figures. For the $K_{S}^{0}$ spectra, except for the last two points at 0.9 TeV and the last three points at 2.76 TeV, all the other points agree with the CSP fit within $30\%$. For the $\rm \Lambda$ spectra, except for the points with $z=$ 6.4 and 8.2 GeV/c at 0.9 TeV, all the other data points are consistent with the CSP fit within $30\%$. For the $\rm \Xi$ spectra, except for the points at $z=$ 4.1, 4.3 and 6.4 GeV/c at 0.9 TeV, all the other data points agree with the CSP fit within $20\%$. For the $\phi$ spectra, except for the last point at 2.76 TeV, all the other points are consistent with the CSP fit within 30$\%$.

\begin{table}[h]
  \caption{\label{tab:CSP_comb_fit_parameters} $C$, $\gamma$ and $\kappa$ returned by the CSP fits on the combination of the scaled $K_{S}^{0}$, $\rm \Lambda$, $\rm \Xi$ and $\phi$ spectra at 0.9, 2.76  and 7 TeV. The uncertainties quoted originate from the statistical and systematic errors of the data points added in quadrature. The last column shows the reduced $\chi^{2}$s for the fits. }
%\begin{center}
\begin{tabular}{@{}ccccc}
\hline\noalign{\smallskip}
\textrm{\ }&
$C$&
$\gamma$&
$\kappa$&
$\chi^{2}$/dof
\\
\hline
${K}_{S}^{0}$ &(166$\pm$4)$\times 10^{-3}$& 0.89$\pm$0.02&3.04$\pm$0.02&272.50/103\\
$\rm \Lambda$&(197$\pm$5)$\times 10^{-4}$&2.54$\pm$0.07&3.80$\pm$0.04&37.31/74\\
$\rm \Xi$&(155$\pm$4)$\times 10^{-4}$&3.44$\pm$0.11&3.83$\pm$0.05&18.86/55\\
$\phi$&(86$\pm$3)$\times 10^{-4}$&1.94$\pm$0.06&3.09$\pm$0.03&20.27/48\\
\hline
\end{tabular}
%\end{center}
\end{table}

From the above statement, we see that the CSP model can successfully describe the scaling behaviour of the strange particle $p_{\rm T}$ spectra at 0.9, 2.76 and 7 TeV. The reason is as follows. $W(x)$ in eq. (\ref{eq:gamma_function}) and $g(x, p_{\rm T})$ in eq. (\ref{eq:frag_function}) are invariant under the transformation $x \rightarrow x^{\prime} = \lambda x$, $\gamma \rightarrow \gamma^{\prime} = \gamma/\lambda$ and $p_{\rm T} \rightarrow p_{\rm T}^{\prime} = p_{\rm T}/\sqrt{\lambda}$. Here $\lambda=\langle S_{n}/nS_{1} \rangle^{1/2}$, where the average is taken over all the clusters decaying into strange particles \cite{string_perco_model_2}. As a result, the strange particle $p_{\rm T}$ spectra in eq. (\ref{eq:CPS_formula}) are also invariant. This invariance is exactly the scaling behaviour we are looking for. Comparing the $p_{\rm T}^{\prime}$ transformation in the CSP model $p_{\rm T}^{\prime} \rightarrow p_{\rm T}^{\prime}\sqrt{\lambda}$ with the one utilized to search for the scaling behaviour $p_{\rm T}\rightarrow p_{\rm T}/K$, we deduce that the scaling parameter $K$ is proportional to $\langle nS_{1}/S_{n} \rangle^{1/4}$. As the degree of string overlap $nS_{1}/S_{n}$ nonlinearly grows with $\sqrt{s}$ \cite{string_perco_model_1, string_perco_model_3}, the scaling parameter $K$ should also increase with $\sqrt{s}$ in a nonlinear trend. That's indeed what we observed in table \ref{tab:id_particles_a_k_parameters}. Therefore the CSP model can qualitatively explain the scaling behaviour for the $K_{S}^{0}$, $\rm \Lambda$, $\rm \Xi$  and $\phi$ $p_{\rm T}$ spectra separately.

In order to determine the nonlinear trend with which $K$  increases with $\sqrt{s}$, we fit the $K$ values at 0.9, 2.76 and 7 TeV for $K_{S}^{0}$, $\rm \Lambda$, $\rm \Xi$ and $\phi$ in table \ref{tab:id_particles_a_k_parameters} with a function $K=\alpha\textrm{ln}(\sqrt{s})+\beta$, where $\sqrt{s}$ is in TeV, $\alpha$ and $\beta$ are free parameters and $\alpha$ characterizes the rate at which $K$ changes with $\textrm{ln}\sqrt{s}$. In sect. \ref{sec:method}, the scaling parameter $K$ at 2.76 TeV is set to be 1 and it is not assigned to an uncertainty. Here, in order to do the fit, we take its uncertainty as the relative error of $\langle p_{\rm T}\rangle$ at this energy. The $\alpha$ values returned  by the fits for $K_{S}^{0}$, $\rm \Lambda$, $\rm \Xi$  and $\phi$ are 0.109$\pm$0.024, 0.120$\pm$0.028, 0.131$\pm$0.031 and 0.085$\pm$0.026. They are consistent within uncertainties. This can be explained by the CSP model as follows. The values of $\langle z \rangle$ are the same for $K_{S}^{0}$ ($\rm \Lambda$, $\rm \Xi$ or $\phi$) at 0.9, 2.76 and 7 TeV. As $K= \langle p_{\rm T}\rangle/\langle z \rangle$, the ratio between the values of $K$ should be equal to the ratio between the values of $\langle p_{\rm T}\rangle$. $\langle p_{\rm T}\rangle$ is evaluated in terms of the CSP model as \cite{pi_k_p_scaling}
\begin{eqnarray}
\langle p_{\rm T}\rangle=\frac{\int_{0}^{\infty}\int_{0}^{\infty}W(x)g(x, p_{\rm T})p_{\rm T}^{2}dxdp_{\rm T}}{\int_{0}^{\infty}\int_{0}^{\infty}W(x)g(x, p_{\rm T})p_{\rm T}dxdp_{\rm T}}.
\label{eq:p_T_CSP}
\end{eqnarray}
Plugging $W(x)$ in eq. (\ref{eq:gamma_function}) and $g(x, p_{\rm T})$ in eq. (\ref{eq:frag_function}) into eq. (\ref{eq:p_T_CSP}), we get
\begin{eqnarray}
\langle p_{\rm T}\rangle=\frac{\sqrt{\gamma\pi}(\kappa-1)\Gamma(\kappa-\frac{3}{2})}{2\Gamma(\kappa)},
\label{eq:mean_pt}
\end{eqnarray}
which depends on $\gamma$ and $\kappa$. In order to determine the values of $\gamma$ and $\kappa$ at 0.9, 2.76 and 7 TeV, we fit the strange particle spectra at these three energies to eq. (\ref{eq:CPS_formula}) with the least squares method. They are tabulated in table \ref{tab:separate_CSP_fit_parameters}.  With these $\gamma$ and $\kappa$ values, we can calculate the ratios between the values of $\langle p_{\rm T}\rangle$ at 0.9 (7) and 2.76 TeV for $K_{S}^{0}$, $\rm \Lambda$, $\rm \Xi$ and $\phi$. They are $0.93\pm0.03$, $0.87\pm0.04$, $0.86\pm0.04$ and $0.95\pm0.38$ ($1.14\pm0.03$, $1.08\pm0.04$, $1.08\pm0.04$ and $1.03\pm0.03$), where uncertainties are due to the errors of $\gamma$ and $\kappa$ at 0.9 (7) and 2.76 TeV. Comparing these ratios with the scaling parameters $K$ at 0.9 and 7 TeV in table \ref{tab:id_particles_a_k_parameters}, we find they are indeed consistent within uncertainties. Therefore, the CSP model can also explain the scaling behaviour of the $K_{S}^{0}$, $\rm \Lambda$, $\rm \Xi$ and $\phi$ $p_{\rm T}$ spectra in a quantitative way.

Finally, we would like to see whether the energy dependence of the scaling parameter $K$ for the strange particles $K_{S}^{0}$, $\rm \Lambda$, $\rm \Xi$ and $\phi$ is the same as that for charged pions, kaons and protons. We fit $K=\alpha\textrm{ln}(\sqrt{s})+\beta$ to the $K$ values at 0.9, 2.76 and 7 TeV for charged pions, kaons and protons in ref. \cite{pi_k_p_scaling}.  The values of $\alpha$ for charged pions, kaons and protons are $0.0638\pm0.0008$, $0.085\pm0.015$ and $0.088\pm0.003$. The $\alpha$ value for charged pions is smaller than those for the strange particles while the $\alpha$ values for charged kaons and protons are comparable to those for the strange particles.

\begin{table}[H]
 \caption{\label{tab:separate_CSP_fit_parameters} $C$, $\gamma$ and $\kappa$ returned by the CSP fits on the ${K}_{S}^{0}$, $\rm \Lambda$, $\rm \Xi$ and $\phi$ spectra at 0.9, 2.76  and 7 TeV. The uncertainties quoted originate from the statistical and systematic errors of the data points added in quadrature. The last column shows the reduced $\chi^{2}$s for the fits.}
\begin{center}
\newsavebox{\tablebox}
\begin{lrbox}{\tablebox}
\begin{tabular}{@{}ccccccc}
\hline
 \textrm{\ }&
$\sqrt{s}$(TeV)&
\textrm{$C$}&
\textrm{$\gamma$}&
\textrm{$\kappa$}&
\textrm{$\chi^{2}$/dof}\\
\hline
\textrm{\ }&0.9 &$(64\pm4)\times 10^{-2}$& 0.83$\pm$0.05&3.13$\pm$0.06&45.08/21\\
${K}_{S}^{0}$&2.76&$(166\pm4)\times 10^{-3}$&0.91$\pm$0.02&3.06$\pm$0.02&70.61/55\\
\textrm{\ }&7&$(79\pm4)\times 10^{-2}$&0.95$\pm$0.05&2.80$\pm$0.03&42.34/21\\
\hline
\textrm{\ }&0.9&$(90\pm4)\times 10^{-3}$&2.14$\pm$0.10&4.02$\pm$0.07&8.55/21\\
$\rm \Lambda$&2.76&$(19\pm2)\times 10^{-3}$&2.57$\pm$0.18&3.80$\pm$0.09&6.53/26\\
\textrm{\ }&7&$(108\pm4)\times 10^{-3}$&2.76$\pm$0.09&3.64$\pm$0.04&6.26/21\\
\hline
\textrm{\ }&0.9&$(72\pm4)\times 10^{-4}$&2.95$\pm$0.27&4.12$\pm$0.18&7.84/19\\
$\rm \Xi$&2.76&$(148\pm3)\times 10^{-5}$&3.52$\pm$0.10&3.82$\pm$0.05&1.37/11\\
\textrm{\ }&7&$(89\pm3)\times 10^{-4}$&3.88$\pm$0.23&3.69$\pm$0.10&3.25/19\\
\hline
\textrm{\ }&0.9&$(152\pm9)\times 10^{-3}$&0.53$\pm$0.41&2.09$\pm$0.44&0.56/1\\
$\phi$&2.76&$(78\pm3)\times 10^{-4}$&2.15$\pm$0.11&3.18$\pm$0.04&7.30/18\\
\textrm{\ }&7&$(102\pm3)\times 10^{-4}$&1.83$\pm$0.06&2.89$\pm$0.03&3.64/23\\
\hline
\end{tabular}
\end{lrbox}
\scalebox{0.85}{\usebox{\tablebox}}
\end{center}
\end{table}

\section{Conclusions}\label{sec:conclusion}

In this paper, we have presented the scaling behaviour of the $K_{S}^{0}$, $\rm \Lambda$, $\rm \Xi$  $p_{\rm T}$ and $\phi$ $p_{\rm T}$ spectra at 0.9, 2.76 and 7 TeV. This scaling behaviour appears when the spectra are shown in terms of the scaling variable $z=p_{\rm T}/K$. The scaling parameter $K$ is determined by the quality factor method and it increases with energy. The rates at which $K$ increases with $\mathrm{ln}\sqrt{s}$ for these strange particles are found to be identical within errors. In the framework of the CSP model, the strange particles are produced through the decay of clusters that are formed by the strings overlapping. We find that the strange mesons and baryons are produced from clusters with different size distributions, while the strange mesons (baryons) $K_{S}^{0}$ and $\phi$ ($\rm \Lambda$ and $\rm \Xi$) originate from clusters with the same size distributions. The cluster's size distributions for strange mesons are more dispersed than those for strange baryons. The scaling behaviour of the $p_{\rm T}$ spectra for these strange particles can be explained by the colour string percolation model quantitatively.
%the dispersions of the cluster's size distribution for strange mesons ($K_{S}^{0}$ and $\phi$) are larger than those for strange baryons ($\rm \Lambda$ and $\rm \Xi$), while the dispersion of the cluster's size distribution for $K_{S}^{0}$ ($\rm \Lambda$) is almost equal to that for $\phi$ ($\rm \Xi$) when considering their errors.  This implies that t

\section*{Acknowledgements}
Liwen Yang, Yanyun Wang, Na Liu, Xiaoling Du and Wenchao Zhang were supported by the Fundamental Research Funds for the Central Universities of China under Grant No. GK201502006, by the Scientific Research Foundation for the Returned Overseas Chinese Scholars, State Education Ministry, by Natural Science Basic Research Plan in Shaanxi Province of China under Grant No. 2017JM1040,  and by the National Natural Science Foundation of China under Grant Nos. 11447024 and 11505108. Wenhui Hao was supported by the National Student's Platform for Innovation and Entrepreneurship Training Program under Grant No. 201710718043.

\end{document}